\documentclass[conference]{IEEEtran}
\ifCLASSINFOpdf
\else
\fi
\ifCLASSOPTIONcompsoc
  \usepackage[caption=false,font=normalsize,labelfont=sf,textfont=sf]{subfig}
\else
  \usepackage[caption=false,font=footnotesize]{subfig}
\fi
\usepackage{amsmath,amssymb}
\usepackage{graphicx}
\usepackage{amsthm}
\theoremstyle{definition}

\newtheorem*{theorem*}{Theorem}

\usepackage{txfonts}
\usepackage{setspace}
\begin{document}
%
\title{Numerical Approach to Maximize SNR\\ for CDMA Systems}

\author{\IEEEauthorblockN{Hirofumi Tsuda}
\IEEEauthorblockA{Department of Applied Mathematics and Physics\\
Graduate School of Informatic\\ Kyoto University\\
Kyoto, Japan\\
Email: tsuda.hirofumi.38u@st.kyoto-u.ac.jp}
\and
\IEEEauthorblockN{Ken Umeno}
\IEEEauthorblockA{Department of Applied Mathematics and Physics\\
Graduate School of Informatic\\ Kyoto University\\
Kyoto, Japan\\
Email: umeno.ken.8z@kyoto-u.ac.jp}}


%


\maketitle

\begin{abstract}
Signal to Noise Ratio (SNR) is an important index for wireless communications. There are many methods for increasing SNR. In CDMA systems, spreading sequences are used. To increase SNR, we have to improve spreading sequences. In classical approaches, the expression of SNR is not differentiable in terms of the parameter of the spreading sequences even in no fading situations. Thus, we express it as the differentiable form and construct the non-linear programing for maximizing SNR. In particular, we solve the problem of maximizing SNR numerically by obtaining spreading sequences whose SNR is guaranteed to be high. 
\end{abstract}


%
\IEEEpeerreviewmaketitle

\section{Introduction}
CDMA is used for the 3G mobile communication system. In CDMA systems, we use spreading sequences as codes to communicate each other in the same band of the frequency. This is a big feature of CDMA systems. We can efficiently use the frequency band with CDMA.\\
Recently, the number of people who communicate each other has been dramatically increasing. However, the band of frequency we can use is limited. The new method to communicate which has high capacity \cite{shannon} is required.\\
In this paper, our topic is about designing direct spreading sequence based CDMA (DS-CDMA) \cite{dscdma}. In general, high Signal to Noise Ratio (SNR) has been demanded for achieving high spectral efficiency \cite{efficiency}. In CDMA, crosscorrelation is a key component of interference noise. It is necessary to reduce crosscorrelation to achieve high capacity. Autocorrelation is related to code synchronization at receiver side. It is desirable that the first peak of crosscorrelation and the second peak in autocorrelation should be kept low. However, there are some limitations about the correlation nature of spreading sequences. Sarwate \cite{sarwate} shows the relationship that there is an avoidable trade-off between lowering crosscorrelation peak and autocorrelation's second peak. This result shows that we cannot have the peak of crosscorrelation and autocorrelation zero. Welch \cite{welch} shows that the lower bounds on the maximal crosscorrelation and autocorrelation is bounded below. \\
The current spreading sequence of 3G CDMA systems is the Gold code \cite{gold}. It is known that this spreading sequence as well as Kasami sequence is optimal in all the binary spreading sequences. In \cite{chaos_cdma} \cite{chaos_mod}, the use of chaotic spreading sequences is proposed. So far, many spreading sequences have been proposed which are obtained from chaotic maps. Examples of such spreading sequences are: based on the Logistic map \cite{logistic}, based on the Chebyshev map \cite{chaos}, based on ergodic theory \cite{ergotic} and the Bernoulli shift \cite{autocor} \cite{mazzini}.\\
On the contrary to those previous works, we derive spreading sequences whose SNR is maximized in a very general setting where we do not assume an explicit mechanism of generating sequence such as maps for chaotic spreading sequence as that in the above works.\\
The expression of the SNR formula is firstly obtained in \cite{mazzini} \cite{pursley}. To increase SNR, we have to reduce the denominator of SNR, namely, variance of interference noise. However, in their expressions, it is difficult to differentiate SNR in terms of the parameter variables of spreading sequences. Here, we show the new SNR formula whose denominator is differentiable in terms of the parameter variables of spreaing sequences. From this expression, we construct the optimization problem: minimize the denominator of SNR. If there are two users in the channel, we obtain the highest SNR to solve the problem. We numerically obtain such solutions dramatically improving SNR which is much greater than existing standard Gold sequences with SLSQP \cite{slsqp} and evaluate these solutions with the Sarwate's limit.

\section{Asynchronous CDMA Model}
We consider the asynchronous CDMA model. We assume no fading signals. This model is taken from \cite{mazzini} \cite{pursley}. Let $N$ be the length of spreading sequences. We assume that the number of users is $K$. The user $k$'s data signal $b_k(t)$ is expressed as
\begin{equation}
b_k(t) = \sum_{n=-\infty}^{\infty} b_{k,n} p_{T}(t - nT),
\end{equation}
where $b_{k,n} \in \{-1,1\}$ is the $n$-th component of bits which user $k$ send, $T$ is the duration of a one symbol and $p_{T}(t)$ is a rectangular pulse defined as
\[p_T(t) = \left\{ \begin{array}{c c}
1 & 0 \leq t \leq T\\
0 & \mbox{otherwise}
\end{array} \right. .\]
The user $k$'s code waveform $s_k(t)$ is expressed as
\begin{equation}
s_k(t) = \sum_{n=-\infty}^{\infty} s_{k,n} p_{T_c}(t - nT_c),
\end{equation}
where $s_{k,n}$ is the $n$-th component of the user $k$'s spreading sequence and $T_c$ is the width of the each chip which satisfies $NT_c = T$. The sequence $(s_{k,n})$ has the period $N$, that is, $s_{k,n}=s_{k,n+N}$. \\
The user $k$'s transmitted signal $\zeta_k(t)$ is
\begin{equation}
\zeta_k(t) = \sqrt{2P} \operatorname{Re}[s_k(t)b_k(t)\exp(j \omega_c t + j\theta_k)],
\label{eq:carrer}
\end{equation}
where $j$ is the unit imaginary number, $P$ is the common signal power, $\omega_c$ is the common carrier frequency and $\theta_k$ is the phase of the $k$-th user.\\
The received signal $r(t)$ is
\begin{equation}
r(t) = \sqrt{2P} \sum_{k=1}^K \operatorname{Re}\left[s_k(t-\tau_k)b_k(t-\tau_k)\exp\left(j \omega_c t + j \psi_k \right) \right] + n(t),
\end{equation}
where $\tau_k$ is the time delay, $\psi_k = \theta_k - \omega_c \tau_k$ and $n(t)$ is the additive white Gaussian noise (AWGN).
If the received signal $r(t)$ is the input to a correlation receiver matched to $\zeta_i(t)$, the corresponding output $Z_i$ is
\begin{equation}
Z_i = \int_{0}^T r(t) \operatorname{Re}[s_i(t-\tau_i)\exp(j \omega_c t + j \psi_i)]dt.
\label{eq:output}
\end{equation}
Without loss of generality, we assume $\tau_i = 0$ and $\theta_i = 0$ and hence $\psi_i = 0$. With a low-pass filter, we can ignore double frequency terms, and transform Eq.(\ref{eq:output}),
\begin{equation}
\begin{split}
Z_i &= \sqrt{\frac{P}{2}} \sum_{k=1}^K \int_{0}^T \operatorname{Re}[s_k(t)b_k(t) \overline{s_i(t)}\exp(j \psi_k)]dt\\
&+ \int_0^T n(t)\operatorname{Re}[s_i(t)\exp(j \omega_c t)],
\label{eq:output2}
\end{split}
\end{equation}
where $\overline{z}$ is the complex conjugate of $z$ and $\displaystyle \overline{s_i(t)} = \sum_{n=-\infty}^{\infty} \overline{s_{i,n}} p_{T_c}(t - nT_c)$. In obtaining  Eq. (\ref{eq:output2}), we used the identity
\begin{equation}
2\operatorname{Re}[z_1]\operatorname{Re}[z_2] = \operatorname{Re}[z_1z_2] + \operatorname{Re}[z_1\overline{z_2}],
\label{eq:thm1}
\end{equation}
where $z_1, z_2 \in \mathbb{C}$.\\
Similar to \cite{pursley}, we assume that the phase $\psi_k$, time delays $\tau_k$ and $b_{k,n}$ are independent random variables and they are uniformly distributed on $[0, 2\pi)$, $[0,T)$ and $\{-1 ,1\}$. Without loss of generality, we assume that $b_{i,0} = +1$.\\
To evaluate SNR, we define
\begin{equation}
\mu_{i,k}(\tau; t) = b_k(t - \tau)s_k(t - \tau)\overline{s_i(t)}
\end{equation}
and
\begin{equation}
\xi_{i, k_1,k_2}(\tau_1, \tau_2; t_1, t_2) = \mu_{i, k_1}(\tau_1; t_1) \overline{\mu_{i, k_2}}(\tau_2; t_2).
\end{equation}
For convenience, we write $\mu_{i,i}$ and $\xi_{i,i,i}$ as $\mu_i$ and $\xi_i$. We divide $Z_i$ into the three signals, the user $i$'s desired signal $D_i$, the interference signal $I_i$ and the AWGN signal $N_i$. They are expressed as
\begin{equation}
\begin{split}
D_i &= \sqrt{\frac{P}{2}} \int_0^T b_i(t)dt\\
I_i &= \sqrt{\frac{P}{2}} \sum_{\substack{k=1 \\ k \neq i}}\operatorname{Re}[\tilde{I}_{i,k}]\\
N_i &= \int_0^Tn(t)\operatorname{Re}[s_i(t)\exp(j\omega_ct)]
\end{split}
\end{equation}
where
\begin{equation*}
\begin{split}
\tilde{I}_{i,k} &= \int_0^T \mu_{i,k}(\tau_k; t) \exp(j \psi_k) dt.\\
\end{split}
\end{equation*}
From these expressions, we decompose $Z_i$ as
\begin{equation}
Z_i = D_i + I_i + N_i.
\end{equation}
Since $\operatorname{E}\{I_i\} = \operatorname{E}\{N_i\} = 0$ and $\displaystyle\operatorname{E}\{D_i\} = T\sqrt{P/2}$, then $\displaystyle \operatorname{E}\{Z_i\} = T\sqrt{P/2}$, where $\operatorname{E}\{X\}$ is the average of $X$.  We assume that $\psi_k$, $\tau_k$ and $b_{k,n}$ are independent. Then, SNR of the user $i$ is defined as
\begin{equation}
\operatorname{SNR}_i = \sqrt{\frac{\operatorname{Var}\{D_i\} }{\operatorname{Var}\{I_i\}  + \operatorname{Var}\{N_i\}}}.
\end{equation}
It is known as in \cite{pursley} that the variance of $N_i$ is
\begin{equation}
\operatorname{Var}\{N_i\} = \frac{1}{4}N_0T
\end{equation}
if $n(t)$ has a two-sided spectral density $\frac{1}{2}N_0$.\\
To derive SNR, we should calculate the variance of $I_i$. The variance of $I_i$ is
\begin{equation}
\operatorname{Var}\{I_i\} =\frac{P}{4}\sum^K_{\substack{k=1 \\ k \neq i}}\operatorname{E}\{|\tilde{I}_{i,k}|^2\}.
\end{equation}
 In the above equation, we used Eq. (\ref{eq:thm1}). It is clear that
\begin{equation}
E_{\psi_k}\left\{\operatorname{Re}\left[\left(\tilde{I}_{i,k}\right)^2\right] \right\} = 0,
\end{equation} 
where $E_{\psi_k}\{X\}$ is the average of $X$ over $\psi_k$. We write the variable over which we take average at the right bottom of $\operatorname{E}$, for example $\operatorname{E}_{\tau_k}\{X\}$.  
The average of $|\tilde{I}_{i,k}|^2$ is expressed as
\begin{equation}
\operatorname{E}\{|\tilde{I}_{i,k}|^2\}= \operatorname{E}_{\mathbf{b}_k,\tau_k}\left\{\int_0^T \int_0^T  \xi_{i,k,k}(\tau_k, \tau_k; t_1,t_2) dt_1dt_2\right\},
\end{equation}
where $\operatorname{E}_{\mathbf{b}_k}\{X\}$ is the average over all the bits of user $k$.
We assume that $l_k T_c \leq \tau_k < (l_k + 1)T_c$, where $0 \leq l_k < N$ is an integer. The double integral term is written as
\begin{equation}
\begin{split}
&\int_0^T \int_0^T  \xi_{i,k,k}(\tau_k, \tau_k; t_1,t_2) dt_1dt_2\\
=& \left|R_{i,k}\left(\tau_k ,0, l_k\right) + \hat{R}_{i,k}\left(\tau_k, 0, l_k\right) \right|^2,
\end{split}
\end{equation}
where
 \begin{equation}
 \begin{split}
 R_{i,k}\left(\tau, n, l \right) =& \left(\tau - nT - lT_c\right)\\
\cdot& \left\{b_{k,-n-1} \sum_{m=1}^{l}\overline{s_{i,m}}s_{k,N-l+m}+ b_{k,-n}\sum_{m=1}^{N-l} \overline{s_{i,l+m}}s_{k,m} \right\},\\
 \hat{R}_{i,k}\left(\tau, n, l\right) =& \left(nT + (l+1)T_c - \tau\right) \left\{b_{k,-n-1} \sum_{m=1}^{l+1}\overline{s_{i,m}}s_{i,N-l+m-1} \right.\\
+&\left. b_{k,-n}\sum_{m=1}^{N-l-1} \overline{s_{i,l+m+1}}s_{k,m} \right\}.
 \end{split}
 \label{eq:cor2}
 \end{equation}
  Note that $R_{i,k}(\tau,n,l)$ and $\hat{R}_{i,k}(\tau,n,l)$ are expressed as the crosscorrelation function in chip-synchronous CDMA systems.
 We define
  \begin{equation}
 \Gamma_{i,k}(\tau,n,l) = \left|R_{i,k}\left(\tau ,n, l\right) + \hat{R}_{i,k}\left(\tau, n, l\right) \right|^2.
   \end{equation}
Therefore, when we take the average over $\tau_k$, $\operatorname{E}\{|\tilde{I}_{i,k}|^2\}$ is
\begin{equation}
\begin{split}
\operatorname{E}\{|\tilde{I}_{i,k}|^2\}&= \frac{1}{T}\cdot \operatorname{E}_{\mathbf{b}_k}\left\{ \int_0^T\int_0^T \int_0^T  \xi_{i,k,k}(\tau_k, \tau_k; t_1,t_2) dt_1dt_2 d\tau_k \right\}\\
&=\frac{1}{T} \cdot \operatorname{E}_{\mathbf{b}_k}\left\{ \sum_{l_k=0}^{N-1}\int_{l_kT_c}^{(l_k+1)T_c}  \Gamma_{i,k}(\tau_k,0,l_k) d\tau_k \right\}.
\end{split}
\end{equation}
Therefore, the variance of $I_i$ is
\begin{equation}
\begin{split}
\operatorname{Var}\{I_{i}\} =& \frac{P}{4T}\sum^K_{\substack{k=1 \\ k \neq i}} \operatorname{E}_{\mathbf{b}_k} \left\{ \sum_{l_k=0}^{N-1} \int_{l_kT_c}^{(l_k+1)T_c} \Gamma_{i,k}\left(\tau_k, 0, l_k\right)  d\tau_k \right\}.
\end{split}
\label{eq:inter_cor}
\end{equation}

\section{New Expression of SNR Formula}
We consider the interference noise term. In \cite{basis}, the crosscorrelation of chip-synchronous CDMA systems can be written in the quadratic form. With this expression, we can transform $\Gamma_{i,k}(\tau_k,0,l_k)$ in Eq. (\ref{eq:inter_cor}) as
 \begin{equation}
  \begin{split}
   &\Gamma_{i,k}(\tau_k,0,l_k) \\
=& \left| \left(\tau_k - l_kT_c\right) \mathbf{s}^*_i B^{(l_k)}_{b_{k,-1},b_{k,0}} \mathbf{s}_k \right. \left.+  \left((l_k+1)T_c - \tau_k\right)\mathbf{s}^*_i B^{(l_k+1)}_{b_{k,-1},b_{k,0}} \mathbf{s}_k \right|^2\\
 =& \left(\tau_k - l_kT_c\right)^2\left|\mathbf{s}^*_i B^{(l_k)}_{b_{k,-1},b_{k,0}} \mathbf{s}_k\right|^2 + \left((l_k+1)T_c - \tau_k\right)^2\left|\mathbf{s}^*_i B^{(l_k+1)}_{b_{k,-1},b_{k,0}} \mathbf{s}_k \right|^2\\
 +&  2\left(\tau_k - l_kT_c\right) \left((l_k+1)T_c - \tau_k\right)\\
 \cdot & \operatorname{Re}\left[\left(\mathbf{s}^*_i B^{(l_k)}_{b_{k,-1},b_{k,0}} \mathbf{s}_k\right) \overline{\left(\mathbf{s}^*_i B^{(l_k+1)}_{b_{k,-1},b_{k,0}} \mathbf{s}_k\right)} \right],
   \end{split}
 \end{equation}
where $\mathbf{z}^*$ is the complex conjugate transpose of $\mathbf{z}$,
\begin{equation}
\mathbf{s}_k = (s_{k,1}, s_{k,2}, \ldots, s_{k,N})^\mathrm{T}
\end{equation}
and
\begin{equation}
B^{(l)}_{b_{k,-1},b_{k,0}} = \left( \begin{array}{c c}
O & b_{k,-1}E_{l} \\
b_{k,0}E_{N-l} & O
\end{array}
\right).
\end{equation}
In the above equation, $\mathbf{s}^\mathrm{T}$ is the transpose of $\mathbf{s}$ and $E_l$ is the identity matrix of size $l$. It can be shown that $\mathbf{s}_k$ is expressed as \cite{basis}
\begin{equation}
\mathbf{s}_k = \frac{1}{\sqrt{N}}\sum_{m=1}^N \alpha^{(k)}_m\mathbf{w}_m(0)=\frac{1}{\sqrt{N}}\sum_{m=1}^N \beta^{(k)}_m\mathbf{w}_m\left(\frac{1}{2N}\right),
\label{eq:basis}
\end{equation}
where $\alpha^{(k)}_m$ and $\beta^{(k)}_m$ are complex coefficients and $\mathbf{w}_m(\eta)$ is the basis vector whose $n$-th component is expressed as
\[\left(\mathbf{w}_m(\eta)\right)_n = \exp\left(2\pi j (n-1)\left(\frac{m}{N} + \eta\right)\right).\]
When we calculate the integral of $\Gamma_{i,k}(\tau_k,0,l_k)$, then,
\begin{equation}
\begin{split}
&\int_{l_kT_c}^{(l_k+1)T_c} \Gamma_{i,k}(\tau_k,0,l_k) d\tau_k\\
 =& \frac{1}{3}T_c^3\left|\mathbf{s}^*_i B^{(l_k)}_{b_{k,-1},b_{k,0}} \mathbf{s}_k\right|^2 + \frac{1}{3}T_c^3\left|\mathbf{s}^*_i B^{(l_k+1)}_{b_{k,-1},b_{k,0}} \mathbf{s}_k \right|^2\\
 +& \frac{1}{3}T_c^3 \operatorname{Re}\left[\left(\mathbf{s}^*_i B^{(l_k)}_{b_{k,-1},b_{k,0}} \mathbf{s}_k\right) \overline{\left(\mathbf{s}^*_i B^{(l_k+1)}_{b_{k,-1},b_{k,0}} \mathbf{s}_k\right)} \right].
\end{split}
\label{eq:inter_term}
\end{equation}
When we take average over the bits $b_{k,-1}$ and $b_{k,0}$, each term of Eq. (\ref{eq:inter_term}) is
\begin{equation}
\begin{split}
\operatorname{E}_{\mathbf{b}_k}\left\{\left|\mathbf{s}^*_i B^{(l_k)}_{b_{k,-1},b_{k,0}} \mathbf{s}_k\right|^2\right\} =& \frac{1}{2}\left\{ \left| \sum_{m=1}^{N}\lambda_m^{(l_k)} \overline{\alpha^{(i)}_m}\alpha^{(k)}_m\right|^2 + \left| \sum_{m=1}^{N}\hat{\lambda}_m^{(l_k)} \overline{\beta^{(i)}_m}\beta^{(k)}_m\right|^2\right\},\\
\end{split}
\end{equation}
and
\begin{equation}
\begin{split}
&\operatorname{E}_{\mathbf{b}_k}\left\{\operatorname{Re}\left[\left(\mathbf{s}^*_i B^{(l_k)}_{b_{k,-1},b_{k,0}} \mathbf{s}_k\right) \overline{\left(\mathbf{s}^*_i B^{(l_k+1)}_{b_{k,-1},b_{k,0}} \mathbf{s}_k\right)} \right]\right\}\\
=&\frac{1}{2} \operatorname{Re}\left[ \left(\sum_{m=1}^{N}\lambda_m^{(l_k)} \overline{\alpha^{(i)}_m}\alpha^{(k)}_m\right) \overline{\left(\sum_{m'=1}^{N}\lambda_m^{(l_k+1)} \overline{\alpha^{(i)}_{m'}}\alpha^{(k)}_{m'}\right)}\right] \\
+& \frac{1}{2} \operatorname{Re}\left[ \left(\sum_{m=1}^{N}\hat{\lambda}_m^{(l_k)} \overline{\beta^{(i)}_m}\beta^{(k)}_m\right) \overline{\left(\sum_{m'=1}^{N}\hat{\lambda}_m^{(l_k+1)} \overline{\beta^{(i)}_{m'}}\beta^{(k)}_{m'}\right)}\right],
\end{split}
\end{equation}
where $\lambda_m^{(l)} =  \exp\left(-2 \pi j l\frac{m}{N}\right)$ and $\hat{\lambda}_m^{(l)} =  \exp\left(-2 \pi j l\left(\frac{m}{N} + \frac{1}{2N}\right)\right)$.
When we take the sum for $l_k$, we arrive at the expression of Eq. (\ref{eq:inter_cor})
\begin{equation}
\operatorname{Var}\{I_{i,k}\} = \frac{PT^2}{12N^2}\sum_{\substack{k=1 \\ k \neq i}}^K \sum_{m=1}^N S^{i,k}_m,
\end{equation}
where
\begin{equation}
\begin{split}
S^{i,k}_m &= \left|\alpha_m^{(i)}\right|^2\left|\alpha_m^{(k)}\right|^2\left(1 + \frac{1}{2}\cos\left(2 \pi \frac{m}{N}\right)\right)\\
&+ \left|\beta_m^{(i)}\right|^2\left|\beta_m^{(k)}\right|^2\left(1 + \frac{1}{2}\cos\left(2 \pi \left(\frac{m}{N} + \frac{1}{2N}\right)\right)\right).
\end{split}
\end{equation}
Then, SNR of the user $i$ is
\begin{equation}
\operatorname{SNR}_i = \left\{\frac{1}{6N^2}\sum_{\substack{k=1 \\ k \neq i}}^K \sum_{m=1}^N S_m^{i,k} + \frac{N_0}{2PT}\right\}^{-1/2}.
\label{eq:SNR}
\end{equation}

\section{Deriving Optimal Spreading Sequences for SNR between Two Users}
We fix the code length $N$ and the number of users $K$. We ignore the Gaussian noise term since it has no relation to spreading sequences. To maximize SNR of the user $i$, we have to minimize the denominator of Eq. (\ref{eq:SNR}). We consider the optimization problem $(\tilde{P})$
\begin{equation}
\begin{split}
(\tilde{P}) & \hspace{3mm} \min \hspace{2mm}\sum_{\substack{k=1 \\ k \neq i}}^K\sum_{m=1}^N S_m^{i,k}\\
\mbox{subject to}  \hspace{3mm} & {\boldsymbol \alpha^{(k)}} = \Phi {\boldsymbol \beta^{(k)}} \hspace{3mm}(k=1,2,\ldots,K),\\
& {\boldsymbol \beta^{(k)}} = \hat{\Phi} {\boldsymbol \alpha^{(k)}} \hspace{3mm}(k=1,2,\ldots,K),\\
&\left\|\boldsymbol \alpha^{(k)}\right\|^2 = \left\|\boldsymbol \beta^{(k)}\right\|^2 = N \hspace{3mm}(k=1,2,\ldots,K),
\end{split}
\end{equation}
where $\|\mathbf{x}\|$ is the Euclidian norm of the vector $\mathbf{x}$,
\begin{equation}
 {\boldsymbol \alpha^{(k)}} = \left( \begin{array}{c}
 \alpha_1^{(k)}\\
 \alpha_2^{(k)}\\
 \vdots\\
 \alpha_N^{(k)}
 \end{array} \right),  {\boldsymbol \beta^{(k)}} = \left( \begin{array}{c}
 \beta_1^{(k)}\\
 \beta_2^{(k)}\\
 \vdots\\
 \beta_N^{(k)}
 \end{array} \right)
  \end{equation}
  and $\Phi$ and $\hat{\Phi}$ are the unitary matrices whose $(m,n)$-th components are
  \begin{equation}
  \begin{split}
   \Phi_{m,n}=&\frac{1}{N}\cdot\frac{2}{1-\exp(2 \pi j (\frac{n-m}{N} + \frac{1}{2N}))},\\
 \hat{\Phi}_{m,n}  =&\frac{1}{N}\cdot\frac{2}{1-\exp(2 \pi j (\frac{n-m}{N} - \frac{1}{2N}))}.
  \end{split}
  \end{equation}
 Note that the objective function is a convex function and this problem is a non-linear convex problem if we remove the norm constraints. The first two constraint terms are obtained from the relation between ${\boldsymbol \alpha^{(k)}}$ and ${\boldsymbol \beta^{(k)}}$. The last constraint terms are signal power constraints. \\
We consider the simplest model that there are two users, that is, $K=2$. In this model, the average of SNR is written as
\begin{equation}
\begin{split}
&\frac{1}{2}\left\{\operatorname{SNR}_1 + \operatorname{SNR}_2 \right\}\\
=& \frac{1}{2}\left[\left\{\frac{1}{6N^2} \sum_{m=1}^N S_m^{1,2} + \frac{N_0}{2PT}\right\}^{-1/2} + \left\{\frac{1}{6N^2} \sum_{m=1}^N S_m^{2,1} + \frac{N_0}{2PT}\right\}^{-1/2}\right]\\
=& \left\{\frac{1}{6N^2} \sum_{m=1}^N S_m^{1,2} + \frac{N_0}{2PT}\right\}^{-1/2} \\
=& \operatorname{SNR}_1.
\end{split}
\end{equation} 
In the above equations, we used the fact that $S_m^{i,k} = S_m^{k,i}$. It is sufficient to consider the problem $(\tilde{P})$ to maximize the average of SNR.\\ 
In general, the variables of an optimization problem are real numbers. However, the variables of the problem $(\tilde{P})$ is complex numbers. We transform  the problem $(\tilde{P})$ to the real number optimization problem. In \cite{cmgc}, it is shown how to transform a complex-number vector to a real-number vector and a complex-number unitary matrix to a real-number orthogonal matrix. Using this method, we consider the problem $(P)$
\begin{equation}
\begin{split}
(P)& \hspace{3mm} \min \hspace{2mm}\sum_{m=1}^N \hat{S}_m^{1,2}\\
\mbox{subject to}  \hspace{3mm} & {\boldsymbol \alpha'^{(k)}} = \Phi' {\boldsymbol \beta'^{(k)}} \hspace{3mm}(k=1,2),\\
& {\boldsymbol \beta'^{(k)}} = \hat{\Phi}' {\boldsymbol \alpha'^{(k)}} \hspace{3mm}(k=1,2),\\
&\left\|\boldsymbol \alpha'^{(k)}\right\|^2 = \left\|\boldsymbol \beta'^{(k)}\right\|^2 = N \hspace{3mm}(k=1,2),
\end{split}
\end{equation}
where
\begin{equation}
\begin{split}
\Phi' &= \left( \begin{array}{c c}
\operatorname{Re}[\Phi] & -\operatorname{Im}[\Phi]\\
\operatorname{Im}[\Phi] & \operatorname{Re}[\Phi]
\end{array} \right), \hat{\Phi}' = \left( \begin{array}{c c}
\operatorname{Re}[\hat{\Phi}] & -\operatorname{Im}[\hat{\Phi}]\\
\operatorname{Im}[\hat{\Phi}] & \operatorname{Re}[\hat{\Phi}]
\end{array} \right), \\
{\boldsymbol \alpha}'^{(k)} &= 
\left( \begin{array}{c}
{\boldsymbol \alpha}^{(k)} _{1}\\
{\boldsymbol \alpha}^{(k)} _{2}
\end{array} \right)
=\left( \begin{array}{c}
\operatorname{Re}[{\boldsymbol \alpha}^{(k)}]\\
\operatorname{Im}[{\boldsymbol \alpha}^{(k)}]
\end{array} \right), {\boldsymbol \beta}'^{(k)}  = 
\left( \begin{array}{c}
{\boldsymbol \beta}^{(k)} _{1}\\
{\boldsymbol \beta}^{(k)} _{2}
\end{array} \right) = 
\left( \begin{array}{c}
\operatorname{Re}[{\boldsymbol \beta}^{(k)}]\\
\operatorname{Im}[{\boldsymbol \beta}^{(k)}]
\end{array} \right)
\end{split}
\end{equation}
and
\begin{equation}
\begin{split}
\hat{S}_m^{i,k}&= \left(\left(\alpha^{(i)}_{1,m}\right)^2 +\left(\alpha^{(i)}_{2,m}\right)^2  \right) \left(\left(\alpha^{(k)}_{1,m}\right)^2 +\left(\alpha^{(k)}_{2,m}\right)^2  \right) \left(1 + \frac{1}{2}\cos\left(2\pi\frac{m}{N}\right)\right)\\
&+\left(\left(\beta^{(i)}_{1,m}\right)^2 +\left(\beta^{(i)}_{2,m}\right)^2  \right)\left(\left(\beta^{(k)}_{1,m}\right)^2 +\left(\beta^{(k)}_{2,m}\right)^2  \right)\\
&\cdot \left(1 + \frac{1}{2}\cos\left(2\pi\left(\frac{m}{N} + \frac{1}{2N}\right)\right)\right).
 \end{split}
\end{equation}
The variables $\alpha^{(k)}_{1,m}$, $\alpha^{(k)}_{2,m}$, $\beta^{(k)}_{1,m}$ and $\beta^{(k)}_{2,m}$ are the $m$-th elements of ${\boldsymbol \alpha}^{(k)}_{1}$, ${\boldsymbol \alpha}^{(k)}_{2}$, ${\boldsymbol \beta}^{(k)}_{1}$ and ${\boldsymbol \beta}^{(k)}_{2}$.
The problem $(P)$ is a non-linear programming. We can obtain the localized solutions with numerical approaches.
\section{Numerical Result} 
To obtain the localized solutions, we use the Sequential Least Squares Programing (SLSQP) \cite{slsqp}. We assume $N_0 = 0$. This algorithm is implemented in the python library, SciPy. We note the language and the libraries which we used in Table \ref{table1}.

\vspace{-10mm}
\begin{table}[htbp]
\centering
\vspace{5mm}
\caption{Language and Library}
\label{table1}
\begin{tabular}{|c | c|}\hline
Language/Library & Version \\\hline\hline
Python & 3.4.3\\\hline
SciPy & 0.15.1\\\hline
NumPy & 1.9.2 \\\hline
\end{tabular}
\end{table}

\begin{figure}[htbp] 
   \centering
   \includegraphics[width=2in]{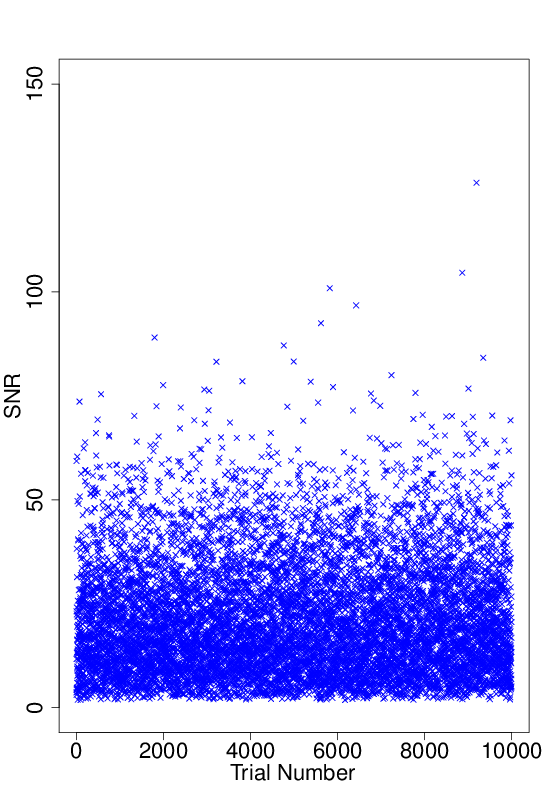} 
   \vspace{-5mm}
   \caption{Numerical calculation of SNR}
   \label{fig:snr}
\end{figure}

We make the random initial point $\boldsymbol \alpha'_0$ and obtain the initial point $\boldsymbol \beta'_0$ from  $\boldsymbol \alpha'_0$. They are expressed as
\begin{equation}
\boldsymbol \alpha'_0 = \left( \begin{array}{c}
\boldsymbol \alpha'^{(1)}\\
\boldsymbol \alpha'^{(2)}\\
\vdots\\
\boldsymbol \alpha'^{(K)}
\end{array} \right), \hspace{2mm} \|\boldsymbol \alpha'^{(k)}\| = N\hspace{2mm}(k=1,2,\ldots,K)
\end{equation}
and
\begin{equation}
\boldsymbol \beta'_0 = \left( \begin{array}{c}
\boldsymbol \beta'^{(1)}\\
\boldsymbol \beta'^{(2)}\\
\vdots\\
\boldsymbol \beta'^{(K)}
\end{array} \right), \boldsymbol \beta'^{(k)} = \hat{\Phi}'  \boldsymbol \alpha'^{(k)}.
\end{equation}
 $\|\boldsymbol\beta'^{(k)}\| = N$ is satisfied if $\| \boldsymbol \alpha'^{(k)}\| = N$ since $\hat{\Phi}'$ is an orthogonal matrix. From these initial points, we obtain a solution with SLSQP. We try to obtain the solution in 10000 times and measure the two types of errors,
\begin{equation}
\begin{split}
e_1 &= \max_{k} \max\left\{\left|N - \|\tilde{\boldsymbol \alpha}'^{(k)}\|^2 \right|, \left|N - \|\tilde{\boldsymbol \beta}'^{(k)}\|^2 \right| \right\},\\
e_2 &= \max_{k} \left\| \tilde{\boldsymbol \beta}'^{(k)} - \hat{\Phi'}\tilde{\boldsymbol \alpha}'^{(k)} \right\|_{\infty},\\
\end{split}
\end{equation}
where $\|\cdot\|_{\infty}$ is $L^{\infty}$ norm. In Table \ref{result}, we note the length $N$, maximum SNR and maximum errors in the trials. It shows that we obtain the solutions which satisfy constraint conditions. 
 Figure \ref{fig:snr} shows the SNR values of the solutions. It shows that the problem $(P)$ has many local solutions.
\begin{table}[htbp]
\centering
\caption{Result}
\label{result}
\begin{tabular}{|c | c|}\hline
Trial Number & 10000 \\\hline
Length $N$ & 31 \\\hline \hline
Maximum SNR & 126.276 \\\hline
Maximum Norm Error $e_1$& 7.461$\times$$10^{-14}$\\\hline
Maximum Matrix Error $e_2$ & 0.0 \\\hline
\end{tabular}
\label{result}
\end{table}

\vspace{-10mm}

\begin{figure}[htbp] 
   \centering
   \includegraphics[width=2.3in]{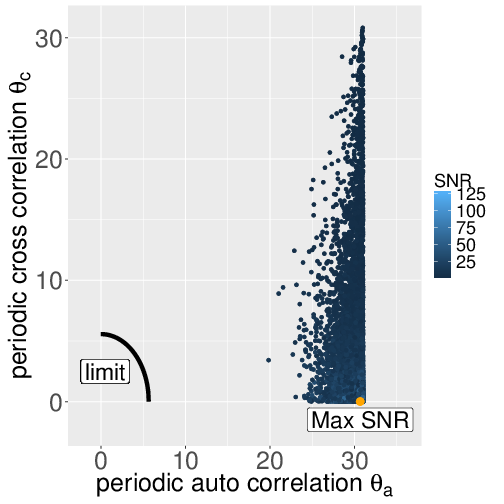} 
   \vspace{-5mm}
   \caption{Relation of periodic correlations : solutions}
   \label{fig:periodic}
\end{figure}

\begin{figure}[htbp] 
   \centering
   \vspace{-3mm}
   \includegraphics[width=2.3in]{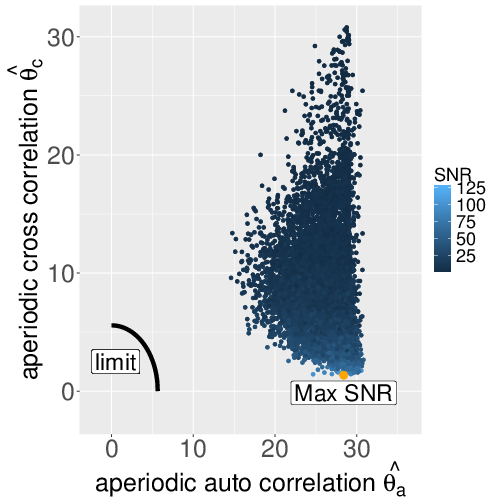} 
   \vspace{-5mm}
   \caption{Relation of aperiodic correlations : solutions}
   \label{fig:aperiodic}
\end{figure}

\begin{figure}[htbp] 
   \centering
   \includegraphics[width=2.3in]{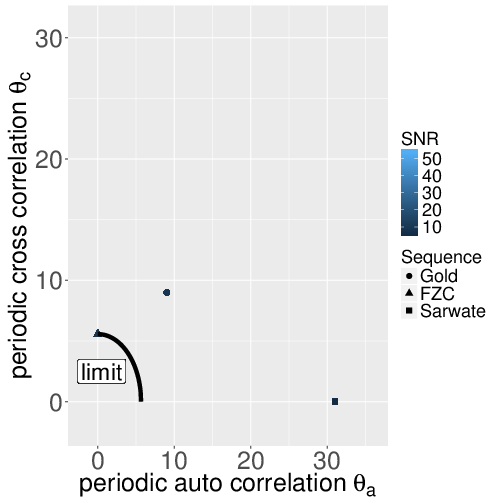} 
   \vspace{-5mm}
   \caption{Relation of periodic correlations : other sequences}
   \label{fig:other_periodic}
\end{figure}

\begin{figure}[htbp] 
   \centering
   \includegraphics[width=2.3in]{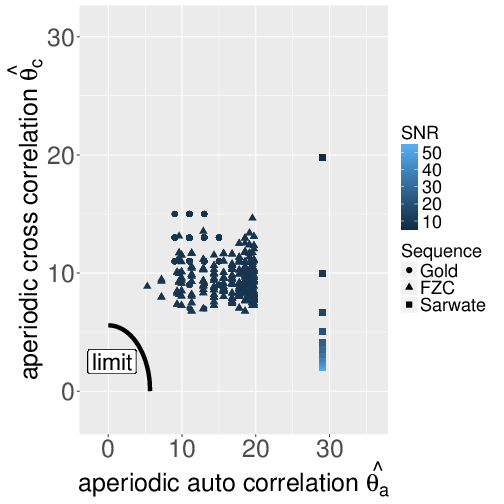} 
   \vspace{-5mm}
   \caption{Relation of aperiodic correlations : other sequences}
   \label{fig:other_aperiodic}
\end{figure}

We evaluate these solutions with the Sarwate's limit \cite{sarwate}. It is expressed as
\begin{equation}
\begin{split}
\left(\frac{\theta^2_c}{N}\right) &+ \frac{N-1}{N(K-1)}\left(\frac{\theta^2_a}{N}\right) \geq 1,\\
\left(\frac{\hat{\theta}^2_c}{N}\right) &+ \frac{N-1}{N(K-1)}\left(\frac{\hat{\theta}^2_a}{N}\right) \geq 1,
\end{split}
\end{equation}
where
\begin{equation}
\begin{split}
\theta_a &= \max \left\{|\theta(u,u)(l)| ; 1 \leq u \leq K, 0 < l \leq N-1 \right\},\\
\theta_c &= \max \left\{ |\theta(u,v)(l)| ; 1 \leq u,v \leq K, u \neq v, 0 \leq l \leq N-1 \right\},\\
\hat{\theta}_a &= \max \left\{|\hat{\theta}(u,u)(l)| ; 1 \leq u \leq K, 0 < l \leq N-1 \right\},\\
\hat{\theta}_c &= \max \left\{|\hat{\theta}(u,v)(l)| ; 1 \leq u,v \leq K, u \neq v, 0 \leq l \leq N-1 \right\}.\\
\end{split}
\end{equation}
In the above equations, $\theta(u,v)(l)$ and $\hat{\theta}(u,v)(l)$ are the periodic correlation function and aperiodic correlation function. From Eq. (\ref{eq:basis}), they are expressed as
\begin{equation}
\begin{split}
\theta(u,v)(l) &= \sum_{m=1}^N \lambda^{(l)}_m \overline{\alpha^{(u)}_m}\alpha^{(v)}_m,\\
\hat{\theta}(u,v)(l) &= \sum_{m=1}^N \hat{\lambda}^{(l)}_m \overline{\beta^{(u)}_m}\beta^{(v)}_m.
\end{split}
\end{equation}
This inequality shows that there is a trade-off between the autocorrelation and the crosscorrelation. We plot the pair of $\theta_a$ and $\theta_c$ and the pair of $\hat{\theta}_a$ and $\hat{\theta}_c$. Figure \ref{fig:periodic} and  \ref{fig:aperiodic} show the result. The bluer point has higher SNR. The orange point is the solution which has the maximum SNR in the trials. From this result, the solutions which have the high SNR have bad autocorrelation. It seems that the pairs of $(\theta_a,\theta_c)$ and $(\hat{\theta}_a,\hat{\theta}_c)$ of solutions would approach to $(N,0)$.\\
We compare the solutions with other sequences. The pair $(\theta_a, \theta_c)$ of the Gold code is well known \cite{gold}. When $N=31$, they are $(\theta_a, \theta_c) = (9,9)$. The Frank-Zadoff-Chu (FZC) sequences \cite{zadoff}\cite{chu} have a good autocorrelation function. The $n$-th element of the FZC sequence, $\mbox{FZC}_n$ is expressed as
 \begin{equation}
\mbox{FZC}_n = \left\{ \begin{array}{c c}
\exp(-j \pi \frac{Mn^2}{N}) & \mbox{if $N$ is even}\\
\exp(-j \pi \frac{Mn(n+1)}{N}) & \mbox{if $N$ is odd}\\
\end{array} \right. ,
\end{equation}
where $M$ is an integer relatively prime to $N$. The pair $(\theta_a, \theta_c)$ of this sequence is $(\theta_a, \theta_c) = (0,\sqrt{N})$ \cite{sarwate}. Sarwate shows the sequences which have a good correlation function. The $n$-th element of it is expressed
 \begin{equation}
 s_n = \exp\left(2 \pi j k\frac{n}{N}\right),
 \end{equation}
 where $0 \leq k \leq N-1$. The pair $(\theta_a, \theta_c)$ of this sequence is $(\theta_a, \theta_c) = (N, 0)$ \cite{sarwate}. We plot the pairs $(\theta_a, \theta_c)$ and $(\hat{\theta}_a, \hat{\theta}_c)$ of these three types sequences in Figure \ref{fig:other_periodic} and \ref{fig:other_aperiodic}. Their SNR is lower than the solution whose SNR is maximal in our trials. It seems to be necessary to obtain the sequence whose $\theta_c$ and $\hat{\theta}_c$ are low to increase SNR.

\section{Conclusion}
In this paper, we showed the new expression of the SNR formula and the numerical approach to obtain the spreading sequences whose SNR is high. With this approach, we can design the spreading sequences without an explicit generation mechanism such as return maps and they are guaranteed to have high SNR.\\
Here, we assume that there is no fading signals and that there are only two users when we solve this SNR maximization problem, such as the first step to solve the optimization problem for maximizing SNR in a very general setting. The remaining issues are to obtain an optimization problem: maximize SNR in a general situation. 
In the next stage, we should analyze the effect of fading signals and more users. 




%

\end{document}